\documentclass[11pt,titlepage]{article}

\usepackage{setspace} 
\usepackage{color}
\usepackage{amsmath}

\usepackage{graphicx}

\usepackage[round,numbers,sort&compress]{natbib} 

\title{Monte Carlo simulations of proteins in cages: influence of confinement 
on the stability of intermediate states}
\author{Pedro~Ojeda~\&~Martin~E.~Garcia\thanks{
           Corresponding author.  Email: magarcia@physik.uni-kassel.de} 
	Theoretische Physik, FB 18,  and Center \\
        for Interdisciplinary Nanostructure Science and Technology (CINSaT), \\
	 Universit\"at Kassel, Germany,
	\and Aurora~Londo\~{n}o \\
	Department of Molecular Biology, Instituto Potosino \\
       de Investigaci\'on Cient\'{\i}fica y Tecnol\'ogica, Camino \\
         a la presa San Jos\'e 2055, 78216 San Luis Potos\'{\i}, Mexico.
	\and Nan-Yow~Chen  \\
          Institute of Physics, Academic Sinica, Nankang, Taiwan
 }

\date{}

\begin{document}

\maketitle
 
\abstract{

We present a theoretical study of the folding of small proteins inside
confining potentials.  Proteins are described in the framework of an
effective potential model which contains the Ramachandran angles as
degrees of freedom and does not need any {\it a priori} information
about the native state. Hydrogen bonds, dipole-dipole- and hydrophobic
interactions are taken explicitly into account. An interesting feature
displayed by this potential is the presence of some intermediates
between the unfolded and native states.  We consider different types
of confining potentials in order to study the structural properties of
proteins folding inside cages with repulsive or attractive
walls. Using the Wang-Landau algorithm we determine the density of
states (DOS) and analyze in detail the thermodynamical properties of
the confined proteins for different sizes of the cages.  We show that
confinement dramatically reduces the phase space available to the
protein and that the presence of intermediate states can be controlled
by varying the properties of the confining potential.  Cages with
strongly attractive walls lead to the disappearance of the
intermediate states and to a two-state folding into a less stable
configuration. However, cages with slightly attractive walls make the
native structure more stable than in the case of pure repulsive
potentials, and the folding process occurs through intermediate
configurations. In order to test the metastable states we analyze the
free energy landscapes as a function of the configurational energy and
of the end-to-end distance as an order parameter.

}
 
\clearpage

\section*{I. INTRODUCTION}

Protein folding is one of the most intensively studied and still
unsolved problems in biology.  Many diseases such as Alzheimer and
Parkinson are believed to be caused by the misfolding and aggregation
of certain proteins~\citep{etienne,kelly,lynn}.  Although in the last
years several aspects related to the Levinthal's paradox have been
clarified with the help of lattice models and other
approaches~\citep{skolnick,abkevich,duan1,go}, many questions
regarding details of the folding and misfolding mechanisms still
remain open.

In this paper we focus on the problem of protein folding assisted  
 by Chaperones, which is one of the mechanisms present in  nature  to
avoid aggregation and misfolding. 
Chaperones are molecules in the form of a cage inside which proteins 
fold correctly. 
 Recently, some progress has been achieved in the
understanding of the folding of  protein inside 
chaperones. These studies have shown that stability and folding
kinetics are strongly correlated with the geometry and the degree of
confinement inside the 
cage~\citep{takagi,thiru,pablo,netto,jewett,fang}. However, many
details of the folding under confinement still remain uncovered.

In this work we focus on the folding of the peptide V3-loop, Protein
Data Bank ID 1NJ0, and analyze it under two kinds of time-independent 
confining potentials. The first 
potential simulates the confining effects of a cage  being
composed by rigid walls, while the second potential describes a cage with 
an attractive inner surface.  
The effect of both potentials are reflected in the
thermodynamical properties, which we calculate using the Wang-Landau
algorithm~\citep{wang,belar}.  

As one of the main results of this work we obtain that 
the folding process of V3-loop occurs through  
metastable intermediate states~\citep{schna}, and that
 the presence of those states can be controlled by the confining potential.

For the description of the protein we use a force field 
 which does not depend on
the previous knowledge of the native structure and is also able to describe 
folding of  proteins into both helices and $\beta$-sheets with the same set
of parameters~\citep{yow}. In addition to this improvement, two new
features not reported previously are included: (i) the dipole-dipole
interaction between the CO-NH pairs lying on the amide plane, and
  (ii) the local hydrophobic interaction between neighboring residues, 
which takes into account the hydrophobic and hydrophilic properties of
the side chains.  The sequence of the amino acids is the only input 
 of the force field. 

The paper is organized as follows. In section II
we describe the model used and  the Monte Carlo
method applied to calculate the thermodynamical
properties of the protein. In section III we present  our results and make
a careful analysis of our simulations. Finally,  we present a summary 
 in section IV.

\section*{II. THEORY}
\subsection*{\label{sec:theory2} The model}

As mentioned in the previous section, the structure of the protein is
simulated using the reduced off-lattice model developed by one of us in 
Ref.~\citep{yow}.  The amino acids are represented by means of 
backbones. Each backbone contains the atoms N, C$_{\alpha}$, C', O and
H. The residues are modeled as spherical beads, $R$, attached to the
C$_{\alpha}$'s. The only remaining degrees of freedom are the
Ramachandran angles $\psi$ and $\phi$. The values for the bond lengths
and angles are given in Ref.~\citep{solo}.

The force field  containing all relevant interactions 
 in the protein is given by 

\begin{equation}
E_{Protein} = E_{Steric}+E_{HB}+E_{DD}+E_{MJ}+E_{LocalHP}
\label{eq:num1}
\end{equation}

\noindent
where $E_{Steric}$ represents hard-core interparticle-potentials to
avoid unphysical contacts, $E_{HB}$ accounts for the hydrogen bonding
and $E_{DD}$ describes the dipole-dipole interactions.  $E_{MJ}$ is a
distance-dependent version of the Miyazawa-Jerningan (MJ)
matrix~\citep{miyaza}, which describes the interactions between
residues.  $E_{LocalHP}$ accounts for local hydrophobic effects. The
role of the presence of water molecules is taken into account both by
the term $E_{MJ}$ and $E_{LocalHP}$.  It is important to point out
that $E_{MJ}$ partially includes the effect of water
polarization~\citep{hao}.  The values of the parameters of this
potential are given in the original work by Chen et al.~\citep{yow}.

In addition to $E_{Protein}$ we
add a term to simulate the confinement of the protein into a cage.
This is accomplished in the present work by using two different kinds of
spherically symmetric potentials depending on a radius $R_c$, which is a measure of the size
of the cage. 
 In a first approach, we use an external potential $V_1$ which allows the protein
to fold freely for distances smaller than $R_c$, but  has a
 strongly repulsive part for larger distances, simulating the presence of the
walls of the cage.  The potential $V_1$ reads~\citep{pablo},

\begin{equation}
V_1= \frac{0.01}{R_c} \left [ e^{r-R_c} (r-1) - \frac{r^2}{2} \right ],
\end{equation} 

\noindent
where $r=|\vec{R}|$ denotes the position of each residue. 
 $V_1$ represents, however,  a too simple description of the confining
potential of a cage.  Therefore, we also investigate the effect of  
  a second external potential $V_2$, which accounts for attractive walls~\citep{lu} and reads 

\begin{eqnarray}
\label{eq:v2}
V_2 &=& 4 \epsilon_{h}\frac{\pi R_c}{r} \left(  \frac{1}{5} \left [  \left ( \frac{\sigma}{r-R_c} 
\right)^{10}  -  \left( \frac{\sigma}{r+R_c} \right)^{10} \right] \right. \nonumber \\
&& - \frac{\epsilon}{2} \left.   \left[  \left( \frac{\sigma}{r-R_c} 
\right)^{4}  -  \left( \frac{\sigma}{r+R_c} \right)^{4} \right]   \right).
\end{eqnarray}

 The physical meaning of the different parameters in Eq.~(\ref{eq:v2})
 can be described as follows.  A uniform distribution of beads spreads
 out on the surface of the cage with a number density
 $1/\sigma^2$. The parameter $\epsilon$ is used to simulate the degree
 of attraction of the inner surface of the cage.  A wall with a purely
 attractive lining has a value of $\epsilon =1$ whereas a purely
 repulsive lining has a value $\epsilon =0$. In Eq.~(\ref{eq:v2}) we
 set $\epsilon_{h}=1.25$ kcal/mol and $\sigma =3.8$~\AA. The external
 potential $V_1$ has the only effect of confining the protein inside
 the cage whereas the external potential $V_2$ interacts with the
 protein by slightly reducing its energy as $\epsilon$ increases. As a
 consequence, the residues tend to be far apart of each other in the
 region close to the walls of the cage.

\subsection*{\label{sec:theory} Simulations}

 Various methods based on Monte Carlo (MC) simulations have been
 proposed to compute the thermodynamical properties of finite systems. They
 include, for instance, multi-canonical simulations~\citep{berg} and
 simulated annealing~\citep{kirk}. In the present work we use the
 Wang-Landau algorithm~\citep{wang}, also including a recent
 improvement introduced by Pereyra et al.~\citep{belar}. One of the
 main advantages of Wang-Landau simulations is that they allow to
 obtain directly the density of states (DOS) of the system, which is,
 of course, independent
 of the simulation temperature. Once the DOS is known, one can obtain
 all the thermodynamical properties of the system at any temperature. 
  Within this
 framework, the transition probability between two conformations
 before and after a MC trial move, $\mathbf{X_1}$ and $\mathbf{X_2}$
 respectively, is calculated as
\begin{equation}  
P ( \mathbf{X_1} \rightarrow \mathbf{X_2} ) = \min \left [ 1, 
\frac{ g ( \mathbf{X_1} )}{ g( \mathbf{X_2} )} \right ], 
\end{equation}

\noindent
where $g(\mathbf{X})$ is the DOS of the system and $\mathbf{X}$ is a
generalized coordinate, which in our case is represented by a vector
with two entries $\mathbf{X}=(E,Q)$, being $E$ the configurational
energy and $Q$ the end-to-end distance. Note that $Q$ can be
interpreted as an order parameter for the folding (unfolding)
transition.

The original scheme developed by Landau \citep{wang} 
 can be briefly described as
follows: one sets the initial function $g(\mathbf{X})$ 
 together with an auxiliary histogram 
$H(\mathbf{X})$ to be equal to 1. Then, each 
time the bin $\mathbf{X}$ is visited, 
one updates the histogram $H(\mathbf{X})$ and modifies 
$g(\mathbf{X})$  as $g(\mathbf{X}) \rightarrow g(\mathbf{X}) \times f$, 
 with $f=e=2.718281...$ .  This procedure is continued 
until a "flat" histogram (with a certain significance,
i.e. $80\% $) is obtained. At this step the histogram $H(\mathbf{X})$ is reseted and the factor 
$f$ is reduced. The usual way to perform this reduction  is by taking
$f_{i+1}=\sqrt{f_i}$.  Convergence is achieved when  a value for 
$f_{i+1}$ close enough to $1$ 
 is obtained. The last step must be  compatible with the desired accuracy, for
example $f=\exp (10^{-7})$.

As mentioned before, we have adopted in this paper a modification
proposed in Ref.~\citep{belar}, which has been demonstrated to speed
up the simulations also to partially avoid the problem of saturation
error.  According to the new scheme, one does not need to wait until
the histogram $H(\mathbf{X})$ is "flat", but it is enough to require
that all the entries of $H(\mathbf{X})$ are visited. Then  
$H(\mathbf{X})=0$ is reseted $f_{i+1}=\sqrt{f_i}$ is updated. 

Following Ref.~\citep{belar} we employ a second histogram
$H_2(\mathbf{X})$ which is never reseted during the whole simulation
and define the Monte Carlo time-step as $t=j/N$, $N$ being the number
of points in the energy axis and $j$ the number of trial moves
performed. If $f_{i+1} \le t^{-1}$ then $f_{i+1} = f(t)= t^{-1}$ and
from this point on $f(t)$ is updated at each Monte Carlo time
step. $H(\mathbf{X})$ is not used during the rest of the calculation.
Convergence is achieved when $f(t)<f_{final}$.  In the present
simulations we used $f_{final}=\exp (10^{-7})$.  Finally, the
thermodynamical properties of the system such as the free energy
$F(T)$, internal energy $U(T)$, entropy $S(T)$ and specific heat
$C(T)$ can be calculated from $g(\mathbf{X})$ as
\begin{equation}
F(T)=-k_B T \ln \left ( \int \mathcal{D} \mathbf{X} g(\mathbf{X}) e^{-\beta E} \right )
\end{equation}
\begin{equation}
{\displaystyle U(T)= \left < E \right > _T = \frac{ {\displaystyle 
\int \mathcal{D} \mathbf{X} E g(\mathbf{X}) e^{-\beta E} } }
{ {\displaystyle \int \mathcal{D} \mathbf{X}  g(\mathbf{X}) e^{-\beta E} } }}
\end{equation}
\begin{equation}
S(T)= \frac{U(T)-F(T)}{T}
\end{equation}
\begin{equation}
C(T)= \frac{ \left < U^2 \right > _T - \left < U \right >^2 _T}{k_B T^2}
\end{equation}

\noindent
where $\beta = 1/k_B T$ and $k_B$ is the Boltzmann constant. 
The free energy landscape as a function of $E$ and $Q$ can be computed as

\begin{equation}
F(E)=-k_B T \ln \left ( \int dQ g(\mathbf{X}) e^{-\beta E} \right ),
\end{equation}

\noindent
and 

\begin{equation}
F(Q)=-k_B T \ln \left ( \int dE g(\mathbf{X}) e^{-\beta E} \right ),
\end{equation}

\noindent
respectively.

\section*{III. RESULTS AND DISCUSSION}

As mentioned in Section I, 
we  focused our attention on a peptide composed of 16 amino acids with PDB code 1NJ0 
to study the folding mediated by confining potentials. 
This peptide conforms the V3-loop of the exterior membrane glycoprotein (GP120) of 
the Human Immunodeficiency Virus type 1 (HIV-1). 

To explore the phase space we have chosen an energy window between
-132.0 kcal/mol and -30 kcal/mol and the end-to-end distance ranging
from 5~\AA~to 50~\AA.  This region in the $\mathbf{X}$ space is enough
to cover both the highly ordered structures present ($T\sim 0$) and
the fully disordered random coils (stable for $T\sim \infty$).  The MC
search was generated by changing each pair of Ramachandran angles
$\psi_i$ and $\phi_i$ at each MC step using cutoffs with values
$|\Delta \psi _c|\le 40^o$ and $|\Delta \phi _c|\le 40^o$. In order to
reach $f_{final}=\exp(10^{-7})$ $8\times 10^9$ trial moves were
necessary.

\begin{figure}[htbp]
\includegraphics[width=7in]{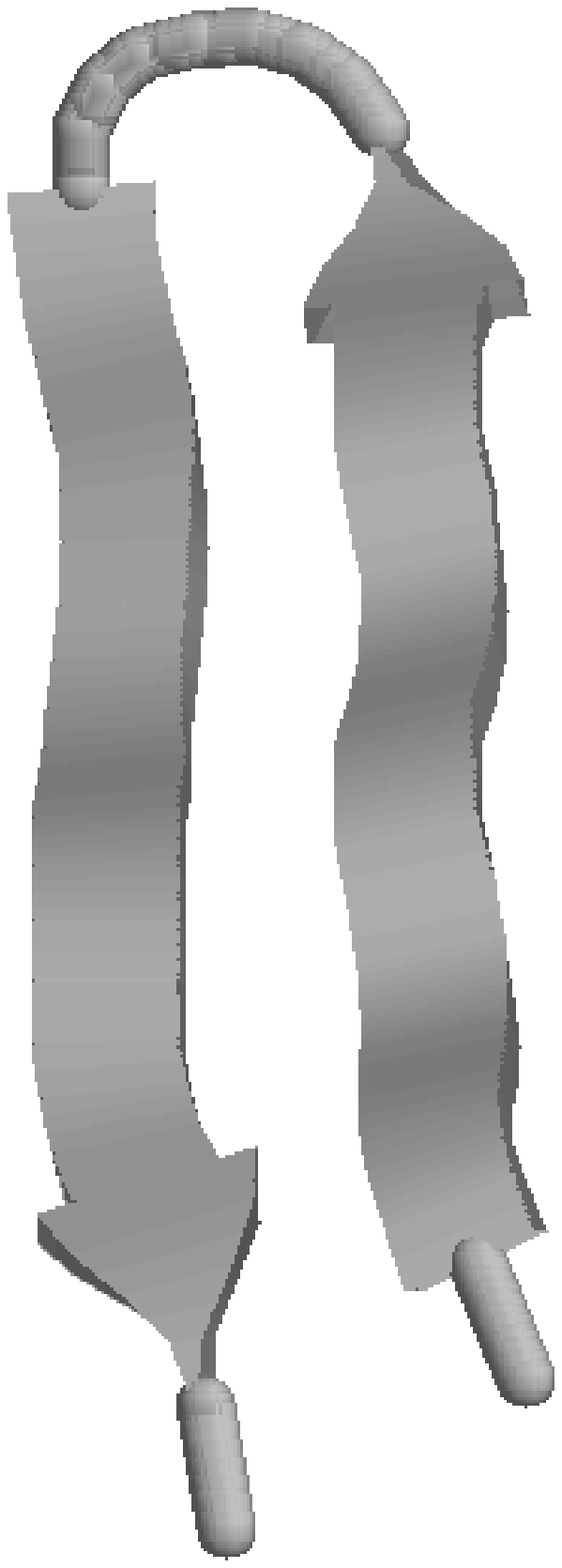}
\caption{}
\label{fig:1njo}   
\end{figure}

The obtained ground state structure of the V3-loop is depicted in
Fig.~\ref{fig:1njo}.  It consists of a $\beta-$sheet structure with
energy $\sim -132.0$ Kcal/mol and an end-to-end distance of $\sim$5.5~\AA.  

A new feature described by  our force field is the presence of intermediates (I) 
between the native (N) and the unfolded (U) states
as shown in Fig.~\ref{fig:free_bulk2}~\citep{schna}.  We obtain two intermediate states 
in the free energy landscape  $F(E)$ at the transition temperature. 
Intermediates are also observed when the energy landscape is
calculated as a  function of the order parameter $Q$ (end-to-end distance).  $F(Q)$  
 is plotted in  Fig.~\ref{fig:free_bulk}.

Interestingly, in $F(Q)$ three intermediates can be clearly observed,
whereas in $F(E)$ only two intermediate states can be
distinguished. This shows the importance of choosing the adequate order
parameters to plot the free energy.
In order to analyze the nature of the intermediate states we have splitted the energetic 
and entropic parts of the free energy. Results indicate that intermediates are mainly 
 due to energy minima and that only the unfolded state is stabilized by entropic effects.

The next problem to address is the influence of confinement on the behavior of the intermediates.
For this purpose,  we calculated the DOS and the specific heat of the protein  
assuming the rigid-wall confining potential $V_1$ described above. We considered different diameters of the cage ($R_c =$ 15~\AA,
20~\AA, and 25~\AA). 
In Fig.~4(a)  we show influence of  $V_1$ on the behavior of the DOS.  
  Note that, due to confinement, $\log [g(E)]$
considerably decreases at high energies (temperatures) compared to the
bulk case ($R_c \rightarrow \infty$).  For energies close to the
ground state, $\log [g(E)]$ does not exhibit any noticeable change because
the protein is almost folded. Since its gyration radius in the ground
state is $R_g \sim 13$~\AA, barriers of radii equal or
larger than 15~\AA~ almost do not affect  folded structures. This
result is consistent with the intuitive picture that cages,  for
instance chaperones, restrict the otherwise huge phase space for high
energies, making the number of available structures considerably
smaller than in absence of a cage.

The effect of confinement can be also observed in the specific heat of
the V3-loop, which we show in Fig.~4(b). There, we plot the
specific heat for different radii of the barriers, 15~\AA, 20~\AA,
25~\AA, and for the bulk case ($R_c\rightarrow \infty$) as function
of $T/T^0_f$, $T^0_f=321$ K being the transition (unfolding) temperature in absence of
a cage.  The effect of the rigid-wall potential $V_1$ is to increase
the transition temperature, see Table~\ref{tab:t2}, and make the curve of the specific heat
broader as the radius of the cage decreases. A broader curve means that
 there are more structures with energies close to the native
state  than in the bulk-case where only the native state
is the most important structure. For radii larger than 25~\AA~
the transition temperatures are equal to $T^0_f$ within the statistical error of our
simulations. 
We conclude that the protein is more stable as the radius of the cage
decreases. This results are in agreement
with Ref. \citep{pablo}, in which   Monte Carlo
simulations were used, and with Refs. \citep{takagi}  and \citep{lu}, where Langevin
 simulations were performed.  It is important to mention, however, that in those cited 
simulations a simplified Go-type potentials was used. 
Thirumalai~\citep{thiru2} made a considerable improvement in the potential by introducing
the effect of the non-native interactions. However,  important interactions
such as dipole-dipole and hydrogen bonds were not taken into account. The 
 presence of intermediates was not reported either in those studies.


Notice that also in the presence of a confining potential $V_1$ the
native state (N) and the unfolded state (U) are separated by two
metastable states (I), as can be seen in Fig.~5.  The largest effect
is observed for a cage with radius 20~\AA, for which the native state
becomes unstable.  In this case, only metastable structures and
unfolded states are present. For all values of $R_c$ we observe the
presence of intermediates. The free energy as a function of the order
parameter $Q$ for the bulk, and barriers of 15~\AA~and 30~\AA~is
showed in Fig.~\ref{fig:free_comp}. The most dramatic changes occur
for the smallest barrier studied (15\AA), for which one of the minima
corresponding to the intermediate states becomes extended.

Now, we introduce the attraction effects of the surface by using the
confining potential $V_2$ (Eq.~\ref{eq:v2} ) with radius $R_c
=30$~\AA. The degree of attraction is described by the coefficient
$\epsilon$.  A completely attractive barrier is obtained when
$\epsilon=1.0$, whereas $\epsilon=0.0$ corresponds to a completely
repulsive or neutral inner wall of the cage 
(potential $V_1$). The effect of $\epsilon$
can be visualized in the following way: as $\epsilon$ increases from 0
to 1, the walls of the cage tend to attract more the residues because
of the relative minimum generated by the potential $V_2$.  The minimum
of $V_2(\epsilon)$ is reached when $\epsilon =1.0$ and corresponds to
$V^{min}_2\sim 5$ Kcal/mol.  This energy is comparable to the energy
required to break one hydrogen bond, $\Delta E_{HB}\sim 4.8$
Kcal/mol. Therefore, for $\epsilon \sim 1.0$ the potential is able to
destroy the structure of the protein (denaturation).  The density of
states for different degrees of attraction and for the bulk case is
shown in Fig. 7(a). The difference
between a purely repulsive barrier and an
attractive one  can be observed in the DOS for energies close to
the ground  state. As $\epsilon$ grows these states increase in number
 with respect to the bulk and determine the folding at the transition
 temperature (see the curves for $\epsilon >0.6$). 

 One can clearly observe a dramatic
reduction of $g(E)$ by $\sim 13$ orders of magnitude as $\epsilon$
goes from 0 to 1. However, this remarkable reduction of the phase
space in this case does not help the protein to fold correctly
but forces it to acquire a denatured conformation. 
  This effect occurs because the peptide decreases its
energy by placing some of the residues close to the border of the
cage. Then, the number of accessible states at those energies
decreases and residues are no longer allowed to be far apart from the
border, since it would cost much energy. As a consequence, the peptide
sticks to the wall of the cage.  As $\epsilon$ increases, the curve of
the specific heat becomes broader (see Fig. 7(b)).  The
transition temperatures for different values of $\epsilon$ and for the
bulk case are presented in Table~\ref{tab:t3}.  
Interestingly, for $\epsilon =0-0.3$
we obtain an increase of the transition temperature compared to
the bulk case.  $\epsilon=0.3$ seems to be the optimal value.  For that
range of $\epsilon$ the protein is more stable than in the absence of a cage.
For higher values of $\epsilon$ the transition temperatures become
lower. For $\epsilon=1.0$ the curve of the specific heat is extremely
broad and attenuated, reflecting the fact that the protein is almost
denatured.

One of the main results of the present
paper is illustrated in Fig.~\ref{fig:free2}, where we plot the free
energy profile $F(E)$ for different values of $\epsilon$ 
and for the bulk (30~\AA~barrier-radius in all cases).  The same native state 
(N), intermediates (I) and unfolded state (U) are observed as long as  $\epsilon
< 0.6$.  However, for $0.6 < \epsilon < 0.8$ the intermediate states (I)
disappear leading to a two-state landscape. This effect is very
important because it shows that one can remove in principle metastable
states by simply changing the electrostatic properties of the
confining surface. 
Shea and coworkers~\citep{jewett} showed that for a particular protein one 
metastable state might exist in the presence  a weakly hydrophobic barrier.  
In this work and for the peptide V3-loop we obtain the opposite result, namely,
 that the protein shows a folding behavior through intermediates in the bulk, but an 
attractive barrier  can lead to eliminate the  intermediate states and to 
induce a two-state folding process. As a function of the order parameter $Q$  the free energy
shows a similar behavior compared to the bulk case but for $\epsilon \sim 0.8$
the intermediates tend to dissappear and two main minima are observed (see Fig.~9).

\section*{IV. CONCLUSION}
 
We have studied the folding of the 16 amino acids peptide 1NJ0 under confining and 
attractive potentials. We have improved previous works by introducing a more 
complete potential which is independent on the native structure and includes 
 relevant interactions such as the dipole-dipole and hydrogen bonds. 
Also, we demonstrate the presence of intermediates not reported before. 
We analyzed the confining effects of a protein inside a Chaperone 
by using two kinds of potentials, one in the form of a rigid-wall 
inert barrier and the other one describing a 
 cage with an attractive inner wall. In the first case we found that 
the presence of the cage tends to decrease the number of accessible states by
allowing only those with are close to the native state. The transition   
temperatures increase as the radius of the barrier decreases as seen in the
curves of the specific heat. These results are in agreement with previous
simulations on other peptides~\citep{takagi,pablo}. Our peptide shows a 
folding through intermediate states~\citep{schna}, which are 
present for any radius of the cage. In the second case  
we considered the effects of attraction inside the barrier ranging
from a completely attractive ($\epsilon=1.0$) to an entirely repulsive
($\epsilon=0.0$) cage wall.
We performed the simulations on a single barrier of radius 30~\AA. 
For fully attractive walls ($\epsilon=1.0$) we observed a decrease of $\sim 13$ orders of
magnitude of the density of states compared to the bulk which can be interpreted
as a denaturation process of the peptide. A strongly attractive potential
$V_2(\epsilon)$ is able to break some hydrogen bonds of the peptide
 as $\epsilon \rightarrow 1.0$, thus decreasing the magnitude of the specific heat 
peak strongly. However, for the interval of $\epsilon$ between 0 and 0.3
we observe that the correct folding of the protein occurs. The increasing transition  
temperatures and the lower average end to end distance 
also allow us to conclude that the protein is more stable as $\epsilon$ increases
in this particular interval. The analysis of the free energy profile
at different values of the parameter $\epsilon$, shows that it is possible to eliminate 
  metastable intermediate states in a controlled way. For the interval
$0.6 < \epsilon < 0.8$ we obtained a two-state folding instead of the
folding through intermediates in the bulk. The simulations were 
performed as a function of the energy $E$ and the end-to-end distance $Q$, which turned out to be an appropriate order parameter.

\medskip

{\small
P. Ojeda thanks the DAAD for the financial support for his PhD. }

\bibliographystyle{biophysj}
\bibliography{article}

\clearpage

\section*{TABLES}

\clearpage

\begin{table}
\begin{tabular}[b]{cc}
\multicolumn{1}{c} {Temperature (K)  } & { Radius (\AA)} \\
\hline
  329.2        &  15       \\
  323.4        &  20      \\
  323.2        &  25      \\
  321.0        &  $\infty$      
\end{tabular}
\caption{Transition temperatures $T_f$ for different values of
the radius $R_c$ of the potential $V_1$ (see main text). Observe that $T_f$ decreases
as the value of the radius increases.}
\label{tab:t2}
\end{table}

\clearpage

\begin{table}
\begin{tabular}[b]{cc}
\multicolumn{1}{c} {Temperature (K)  } & { Radius (\AA)} \\
\hline
  321.0        &  BULK    \\
  324.2        &  $\epsilon =0.0$      \\
  324.1       &   $\epsilon =0.2$    \\
  314.5        &  $\epsilon =0.4$  \\
  292.1      &  $\epsilon =0.6$  \\
  253.5        &  $\epsilon =0.8$  \\
  --        &  $\epsilon =1.0$  \\
\end{tabular}
\caption{Transition temperatures $T_f$ for the confining potential $V_2$ (see main text) for 
different degrees of hydrophobicity, $\epsilon$= 0.0, 0.2, 0.4, 0.6, 0.8, 1.0 
and the bulk case. Notice that in general  $T_f$ decreases
as $\epsilon$  increases. The temperature at $\epsilon=1.0$ is almost ficticious
because the specific heat is almost completely attenuated. }
\label{tab:t3}
\end{table}

\clearpage

\section*{FIGURE LEGENDS}

\clearpage



\subsubsection*{Figure~\ref{fig:1njo}.}
Ground state structure ($\beta$-sheet) of the peptide 1NJ0 ($E_g \sim -132.0$).

\subsubsection*{Figure~\ref{fig:free_bulk2}.}
Free energy as a  function of the configurational energy $E$ for the
bulk ($R_c \rightarrow \infty$) showing the native (N),
intermediates (I) and unfolded (U) states.

\subsubsection*{Figure~\ref{fig:free_bulk}.}
Free energy as a  function of the end-to-end distance $Q$ for the
bulk ($R_c \rightarrow \infty$). We observe the native (N),
unfolded (U) and three intermediate states (I).

\subsubsection*{Figure~\ref{fig:cvst}.}
(a) Logarithm of the density of states 
 $g(E)$ for the potential barrier $V_1$ and for different values of $R_c$, 
15~\AA, 20~\AA, 25~\AA~and for the bulk case. One notices the remarkable decrease of 
the density of states for decreasing $R_c$. 

(b)
Specific heat for the bulk-case and for  barriers with radii 15~\AA, 20~\AA, and 25~\AA.
$T_f= 321$ K is the transition temperature for the bulk. The transition temperature
increases  as the the radius $R_c$ decreases.


\subsubsection*{Figure~\ref{fig:free}.}
Free energy profile $F(E)$ at the transition temperature $T_f$ for the bulk, and barriers
of 15~\AA, 20~\AA~and 25~\AA, showing the folding through intermediates (I) behavior.
The two metastable intermediate states are present in all the cases.

\subsubsection*{Figure~\ref{fig:free_comp}.}
Free energy profile $F(Q)$ at the transition temperature $T_f$ for the bulk, and barriers
of 15~\AA~and 30~\AA, the same features are shown in all these three cases but 
smoothing out is more appreciated for the smallest barrier.


\subsubsection*{Figure~\ref{fig:cvst3}.}
(a) Logarithm of $g(E)$ for different degrees of hydrophobicity $\epsilon=$ 0.0, 0.2, 0.4, 
0.6, 0.8 and 1.0, and for the bulk case. We observed an abrupt decay of $g(E)$
of $\sim 13$ orders of magnitude as $\epsilon$ goes from 0.0 to 1.0. 
For high values of $\epsilon$ the protein tends to be in the unfolded state.

(b) Specific heat of the protein for different values of $\epsilon$, 0.0, 0.2, 0.4, 0.6, 0.8, 1.0,
compared to the bulk case. $T_f= 321$ K is the transition temperature for the bulk.
Notice how the transition temperatures and the peak of the specific heat decrease
as we go from a purely repulsive barrier $\epsilon=0.0$ to a strongly  attractive 
one $\epsilon=1.0$. 



\subsubsection*{Figure~\ref{fig:free2}.}
Free energy profile $F(E)$ at the transition temperature $T_f$ for the bulk, and values of 
$\epsilon$ 0.0, 0.4 and 0.6. Observe how the intermediate states (I) present in the
bulk are destroyed by an attractive surface at $\epsilon =0.6$ causing a two-state folding
process. 

\subsubsection*{Figure~\ref{fig:free_ener_h}.}
Free energy profile $F(Q)$ at the transition temperature $T_f$ for the bulk, and values of 
$\epsilon$ 0.0, 0.4 and 0.8. For small values of $\epsilon$ the free energy is similar
to the bulk but for high values of $\epsilon$ the intermediates tend to dissappear. 
%



\clearpage

\clearpage

\begin{figure}[htbp] 
\centering
\includegraphics[width=7in,angle=90]{free_bulk2.eps}
\caption{}
\label{fig:free_bulk2}
\end{figure} 

\clearpage

\begin{figure}[htbp] 
\centering
\includegraphics[width=7in,angle=90]{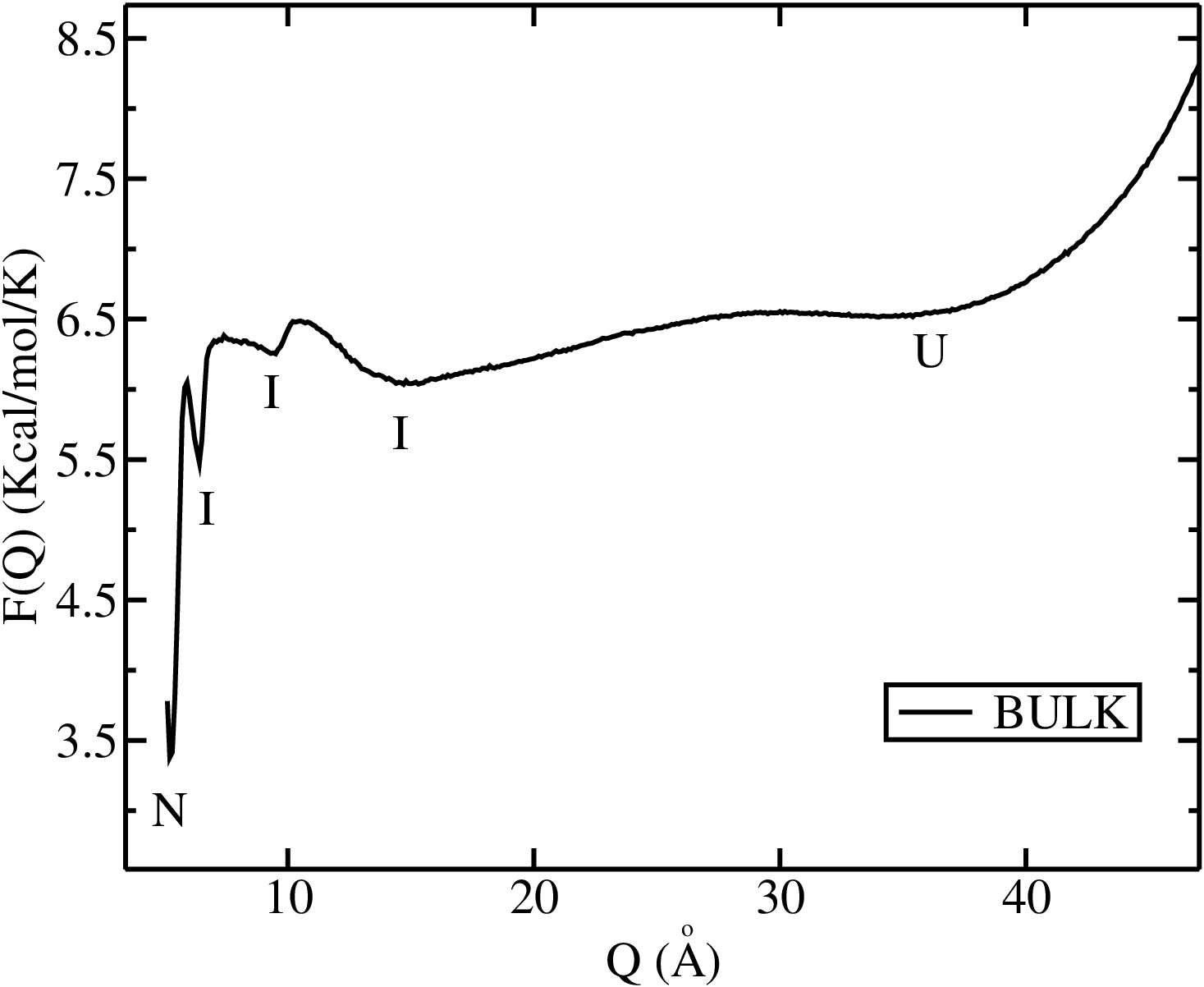}
\caption{}
\label{fig:free_bulk}
\end{figure}


\clearpage

\begin{figure}[htbp] 
\includegraphics[width=7in,angle=90]{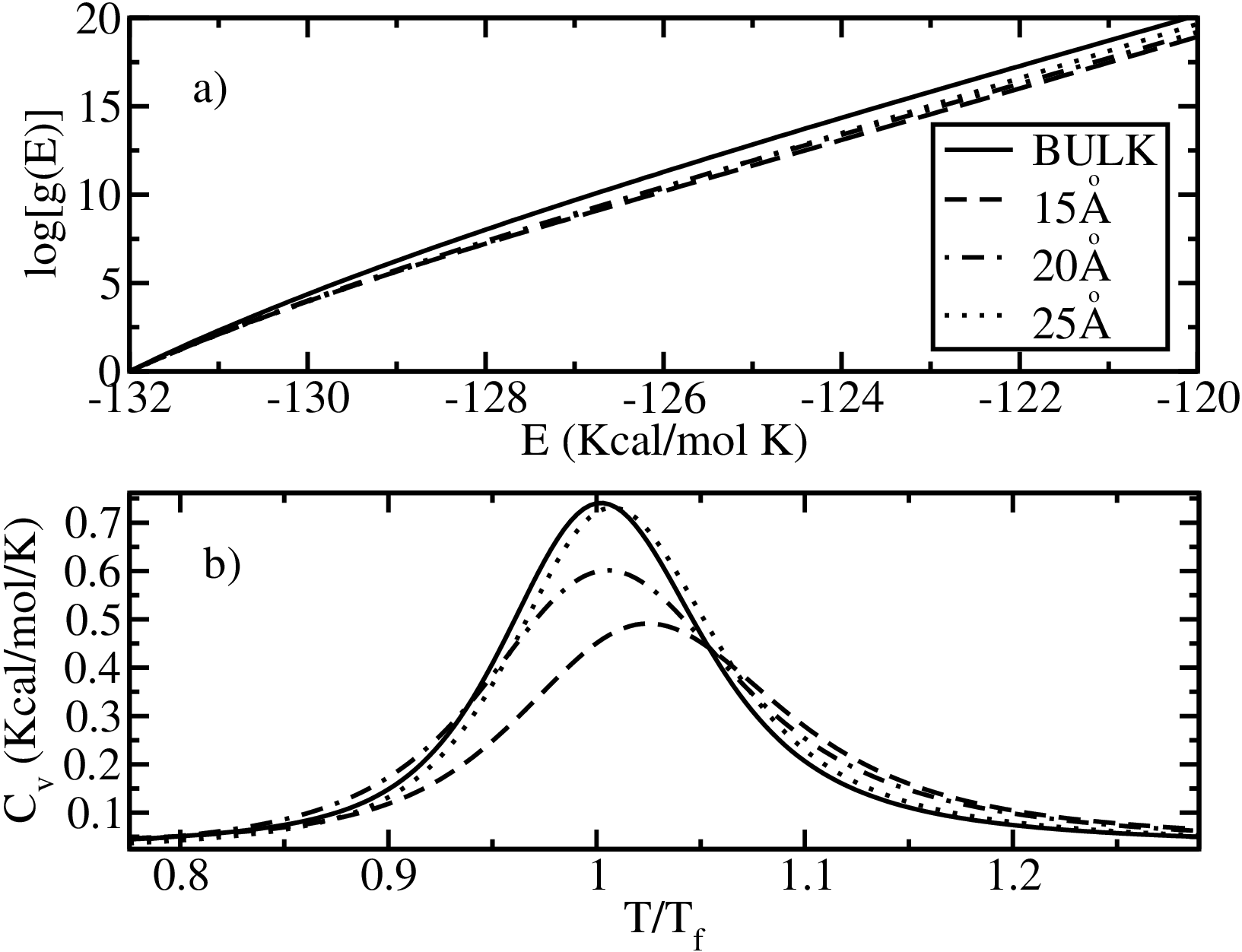}
\caption{}
\label{fig:cvst} 
\end{figure}

%

\clearpage

\begin{figure}[htbp] 
\includegraphics[width=7in,angle=90]{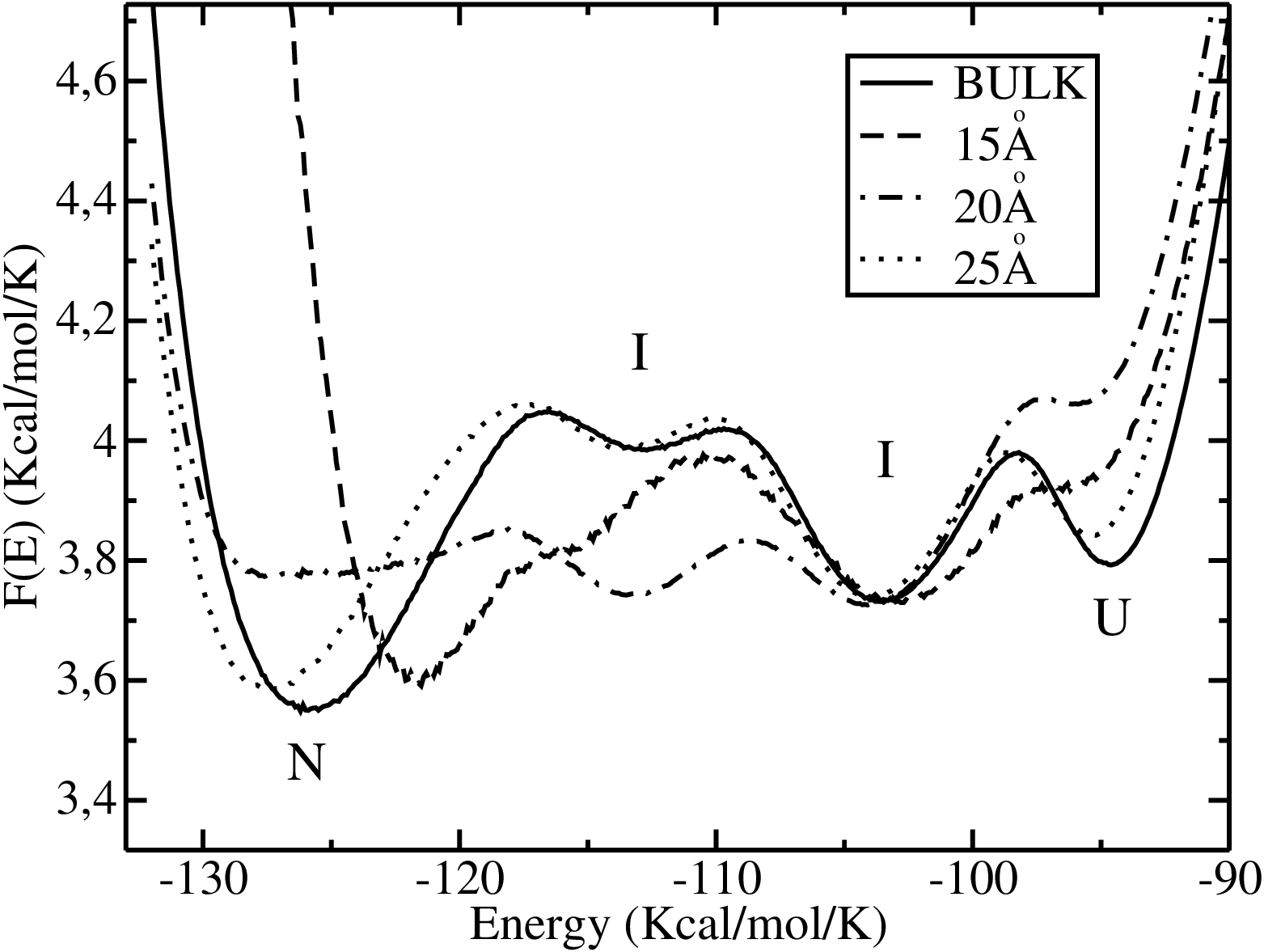}
\caption{}
\label{fig:free} 
\end{figure}

\clearpage

\begin{figure}[htbp] 
\includegraphics[width=7in,angle=90]{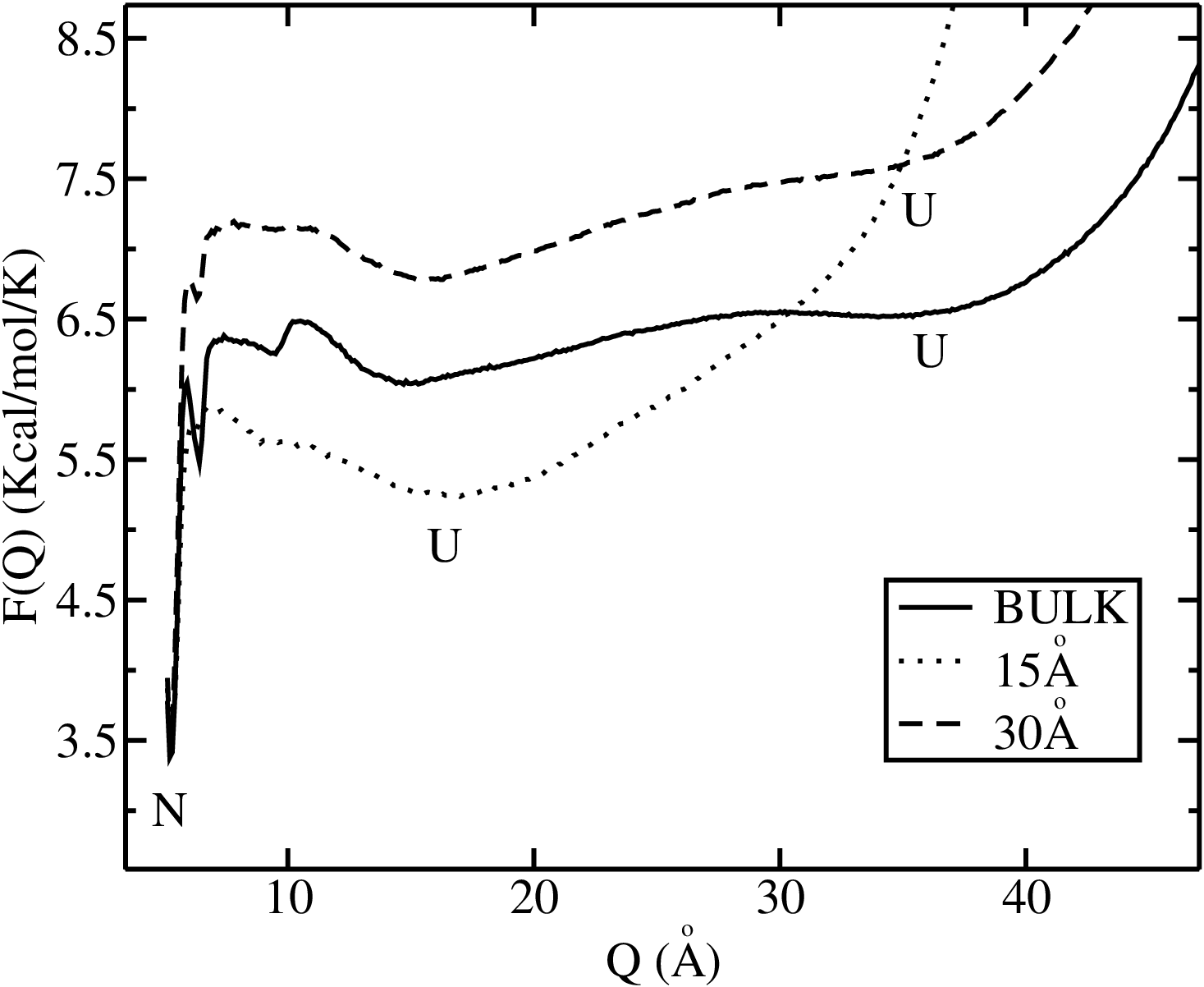}
\caption{}
\label{fig:free_comp} 
\end{figure}



\clearpage


\clearpage

\begin{figure}[htbp]
\includegraphics[width=7in,angle=90]{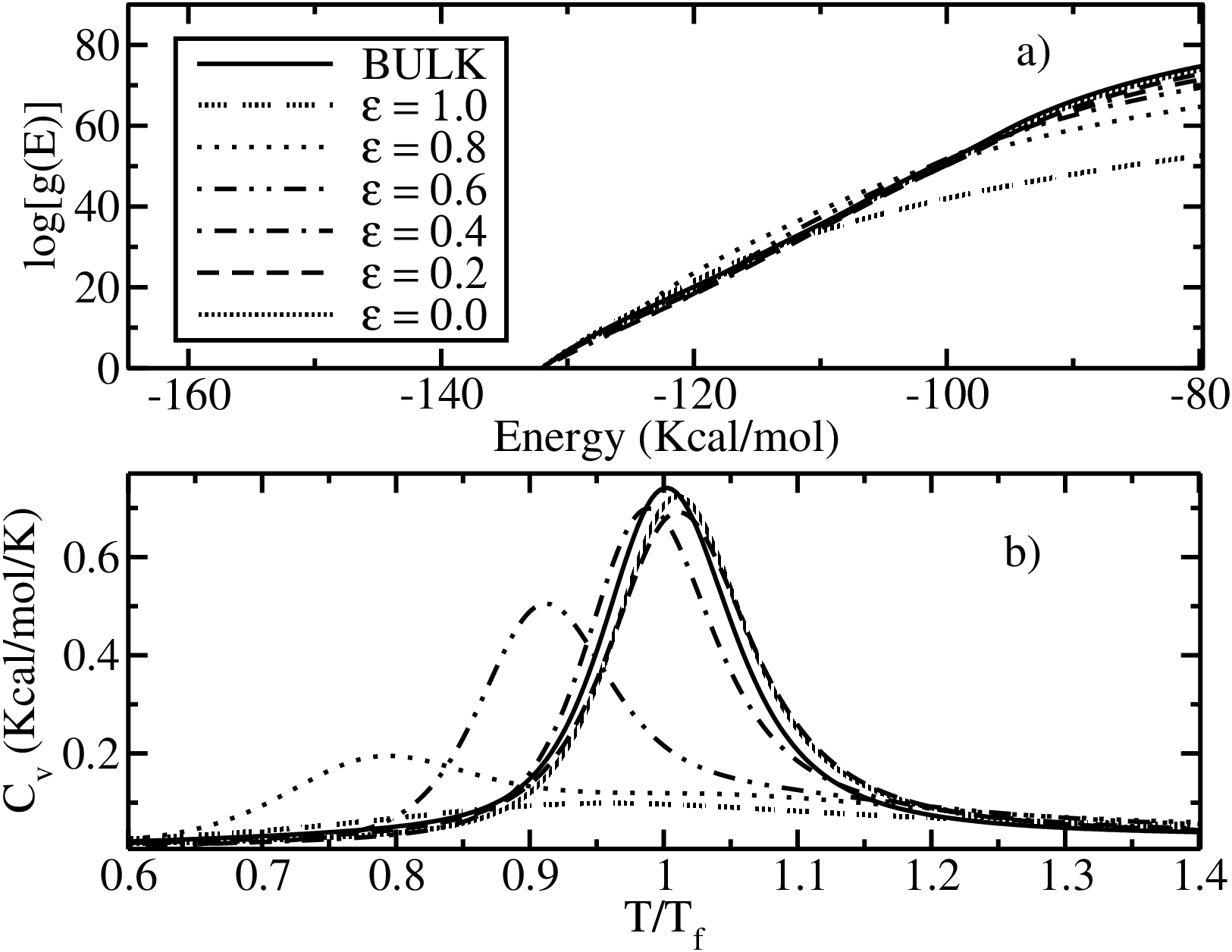}
\caption{}
\label{fig:cvst3}
\end{figure}



\clearpage

\begin{figure}[htbp]
\includegraphics[width=7in,angle=90]{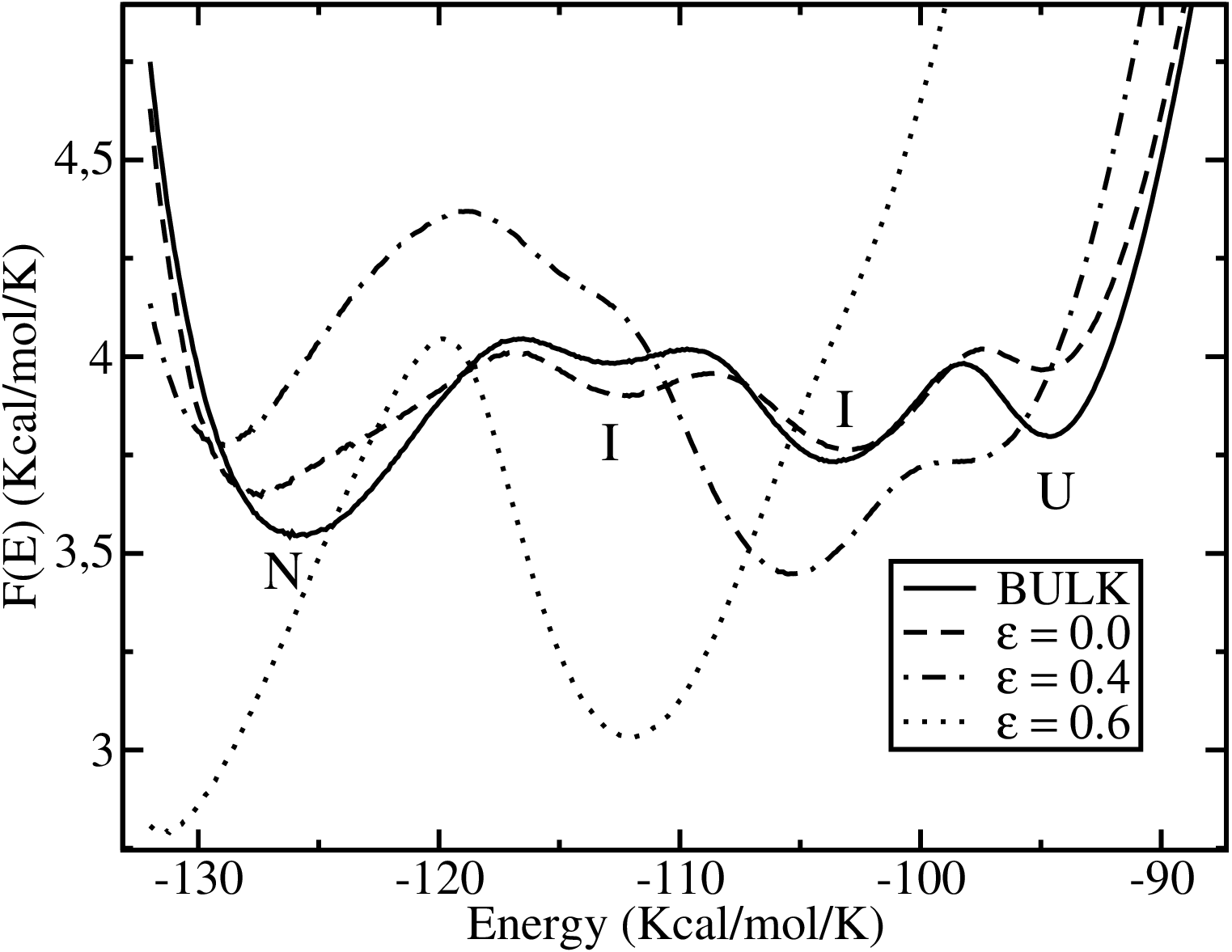}
\caption{}
\label{fig:free2}
\end{figure}

\clearpage

\begin{figure}[htbp]
\includegraphics[width=7in,angle=90]{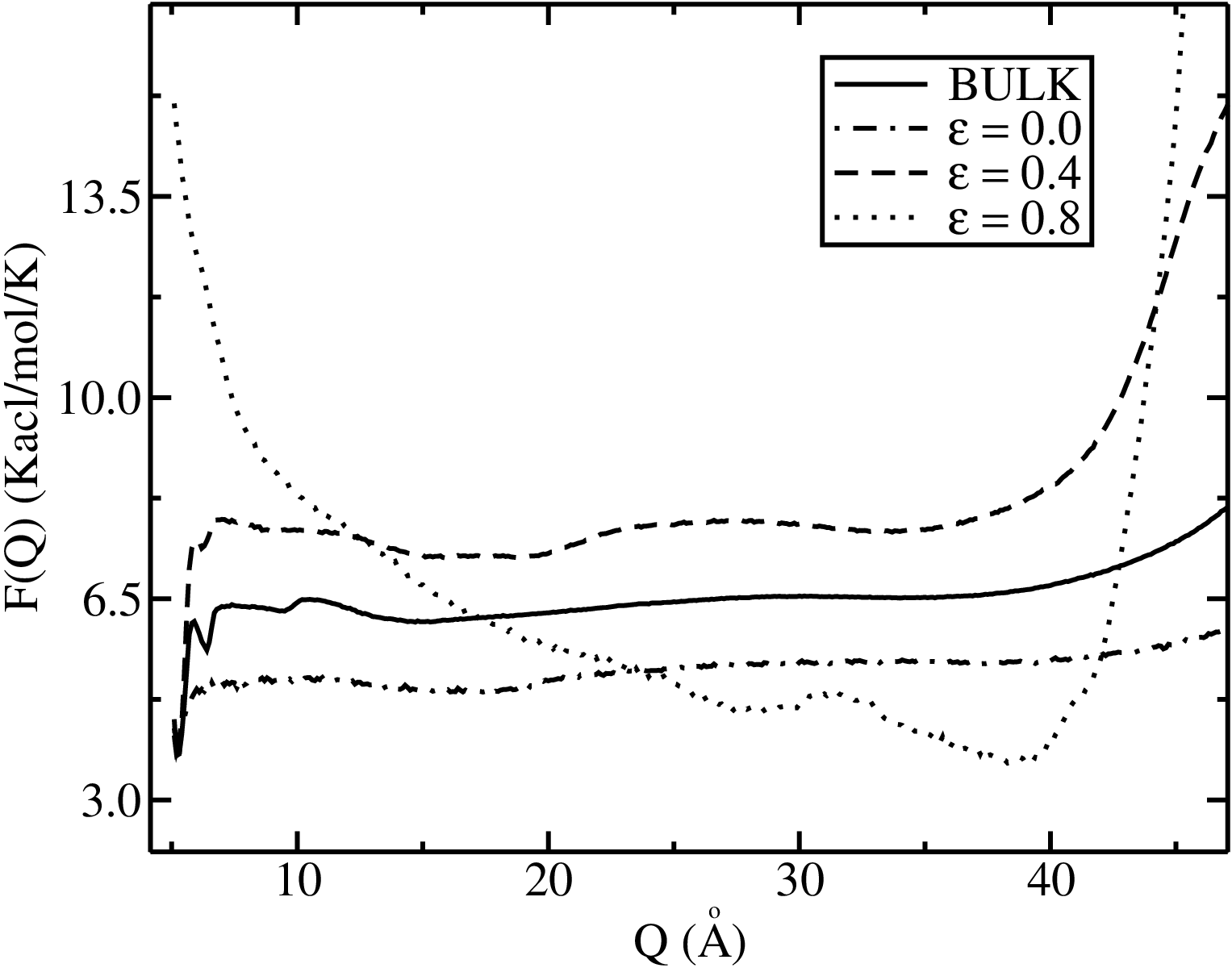}
\caption{}
\label{fig:free_ener_h}
\end{figure}

\end{document}